\documentclass[aps,prb,twocolumn]{revtex4}
\usepackage{graphicx}
\begin{document}
\title{Multi-photon structures in the sub-cyclotron-frequency
range in microwave photoresistance of a two-dimensional electron system}
\author{X.L. Lei and S.Y. Liu}
\affiliation{Department of Physics, Shanghai Jiaotong University,
1954 Huashan Road, Shanghai 200030, China}
\date{\today}

\begin{abstract}

The frequency dependence of the peak-valley pairs occurring 
in the magnetoresistivity of a two-dimensional 
electron system under enhanced microwave irradiation, which are 
considered to associate with multiphoton processes,  
is examined in the sub-cyclotron-frequency range, based on a theoretical 
treatment with photon-assisted electron transitions due to impurity scattering.
It is shown that with equivalent radiation power (producing the same height 
of the main oscillation peak), 
much more and stronger multi-photon structures show up at lower frequency, 
and when frequency increases all these structures 
rapidly weaken, diminish and finally disappear completely.
These are in agreement with the recent experimental observation [cond-mat/0608633].

\end{abstract}


\maketitle

The microwave-induced magnetoresistance oscillations (MIMOs)
in high-mobility two-dimensional (2D) electron systems\cite{Ryz,Zud01,Ye,Mani,Zud03,Dor03} 
continue to be a phenomenon of great interest.
In addition to well-established main oscillations featuring large maximum-minimum
pairs, secondary peak-valley structures were also observed in the early 
experiments\cite{Mani,Zud03,Dor03},
and predicted theoretically\cite{Durst,Lei03}, referred to the effect 
of two- and three-photon processes. Later experimental and theoretical
investigations with enhanced radiation intensity or reduced radiation frequency
disclosed further details of these structures.\cite{Will,Zud04,Mani04,Dor05,
Zud05,Lei05,Lei06apl} 

Motivated by a recent measurement of MIMOs at the subharmonics of 
cyclotron resonance,\cite{Dor0608633} we performed further examination 
using the balance-equation approach\cite{Lei03,Lei05} 
with photon-assisted electron transitions 
due to impurity scattering. Some results obtained by taking the same material 
parameters as those in Ref.\,[\onlinecite{Lei06apl}], are presented here.

\begin{figure}
\includegraphics [width=0.43\textwidth,clip=on] {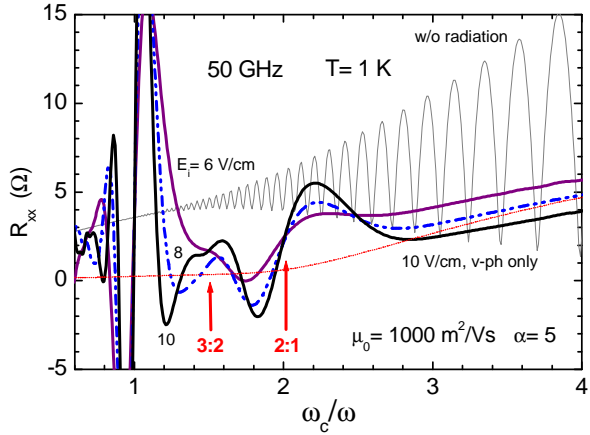}
\vspace*{-0.2cm}
\caption{The magnetoresistivity $R_{xx}$ of a GaAs-based 2DEG 
with $N_{\rm e}=3.0\times 10^{15}$\,m$^{-2}$, $\mu_0=2000$\,m$^2$/Vs and $\alpha=5$, 
subjected to 50\,GHz radiations of incident amplitudes $E_{{\rm i}s}=6,8$ and 
$10$\,V/cm at lattice temperature $T=1$\,K.}
\label{fig1}
\end{figure}
\begin{figure}
\includegraphics [width=0.43\textwidth,clip=on] {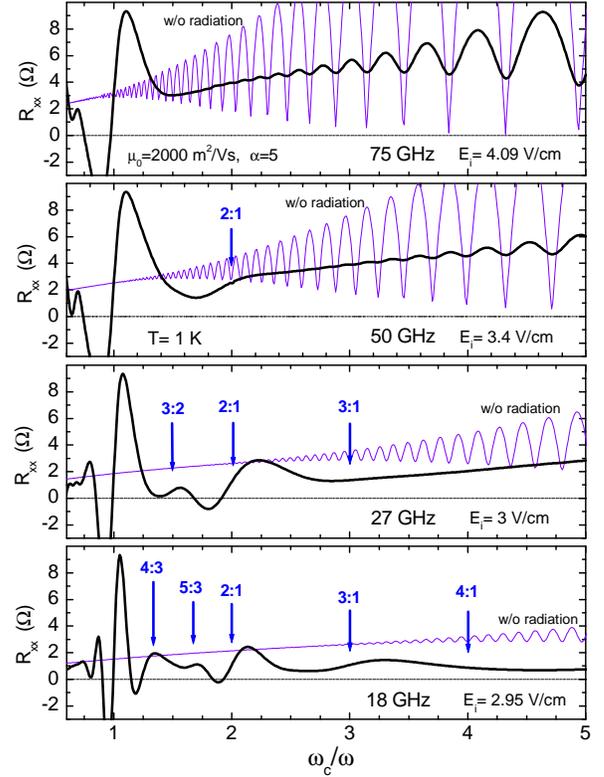}
\vspace*{-0.2cm}
\caption{The Magnetoresistivity $R_{xx}$ of a GaAs-based 2DEG 
with $N_{\rm e}=3.0\times 10^{15}$\,m$^{-2}$, $\mu_0=2000$\,m$^2$/Vs and $\alpha=5$, 
subjected to radiations of frequencies $\omega/2\pi=18,27,50$ and 75\,GHz with
incident amplitudes $E_{\rm i}=2.95, 3, 3.4$ and 4.09\,V/cm respectively,
at lattice temperature $T=1$\,K.} 
\label{fig2}
\end{figure}

Figure 1 shows the calculated magnetoresistivity $R_{xx}$ 
as a function of $\omega_c/\omega$ ($\omega_c$ is the cylotron frequency)
for a GaAs-based 2D system 
having electron density 
$N_{\rm e}=3.0\times 10^{15}$\,m$^{-2}$, 
linear mobility $\mu_0=2000$\,m$^2$/Vs and broadening parameter 
$\alpha=5$, irradiated by microwaves of frequency $\omega/2\pi=50$\,GHz 
having three incident amplitudes $E_{\rm i}=6, 8$ and 10\,V/cm 
at lattice temperature $T=1$\,K.  
Prominent valley-peak pairs show up around $\omega_c/\omega=2$ and 1.5
for all three radiation strengths, and $\omega_c/\omega=2$ appears to be a 
node point with almost zero resistivity response to these
amplitude change.   

Figure 2 illustrates the effect of different radiation frequencies 
on these fine structures associated with multiphoton assisted processes 
for the same system as described in Fig.\,1. 
The radiation strength at each frequency is so chosen that 
almost the same height of the main oscillation peak (around $\omega_c/\omega=1$) 
is obtained for all four frequencies.
However, the valley-peak structures around $\omega_c/\omega=2, 3/2, 3, 4/3$
and 4, which show up prominently at $18$\,GHz, weaken or diminish 
at $27$\,GHz, become barely appreciable only around $\omega_c/\omega=2$ at $50$\,GHz, 
and completely disappeared at 75\,GHz.    

It is also seen from both figures that the average of 
all the resistivity curves with strong irradiation drop down 
below the that of dark curve in the 
sub-cyclotron-frequency range .

\end{document}